\documentclass[genTeX]{nrc1}
\usepackage[pctex32]{graphicx}
\def\be{\begin{equation}}
\def\ee{\end{equation}}
\def\bea{\begin{eqnarray}}
\def\eea{\end{eqnarray}}

\volyear{xx}{2005(?)}
\journal{Can. J. Phys.}
\received{Sept. 10, 2005}
\accepted{?????}
\begin{document}

\title{Looking Beyond Inflationary Cosmology}
\author[Robert H. Brandenberger]{Robert H. Brandenberger}
\address{Department of Physics, McGill University, Montreal, QC 
H3A 2T8, Canada. \email{rhb@hep.physics.mcgill.ca}}

\shortauthor{Brandenberger}

\maketitle
\begin{abstract}
In spite of the phenomenological successes of the inflationary universe
scenario, the current realizations of inflation making use of scalar
fields lead to serious conceptual problems which are reviewed in this
lecture. String theory may provide an avenue towards addressing these
problems. One particular approach to combining string theory and
cosmology is String Gas Cosmology. The basic principles of this
approach are summarized.
\\\\PACS Nos.:  98.80.Cq
\end{abstract}
\begin{resume}
Malgr\'e les succes ph\'enom\'enologiques du scenario de l'univers 
inflationnaire, les impl\'ementations actuelles du mod\`ele
inflationnaire en utilisant des champs scalaries m\`enent \`a des
probl\`emes conceptuels, la discussion desquels forme la
premi\`ere partie de cette conf\'erence. La th\'eorie des
supercordes engendre la possibilit\'e de r\'esoudre ces probl\`emes.
``String Gas Cosmology'' pr\'esente un chemin int\'eressant pour combiner la 
th\'eorie des supercordes et la cosmologie. Un sommaire des aspects de base 
de ce mod\`ele est pr\'esent\'e dans la deuxi\`eme partie de cette
conf\'erence. 

\end{resume}

\def\tablefootnote#1{%
\hbox to \textwidth{\hss\vbox{\hsize\captionwidth\footnotesize#1}\hss}} 

\section{Introduction}

The inflationary universe scenario is the current paradigm for understanding
the evolution of the very early universe. Developed about 25 years ago
\cite{Guth,Sato} (see also \cite{Starob1,Brout} for earlier but
related work), inflationary cosmology has been extremely successful
from a phenomenological point of view. It predicted the spatial flatness
of the universe, and - most importantly - provided the first theory
for the origin of the large-scale structure of the universe based on
fundamental physics \cite{ChibMukh,Lukash} (see also \cite{Sato,Press}
for qualitative arguments and \cite{Starob2} for an analysis of the
predicted spectrum of gravitational waves based on the model of 
\cite{Starob1}). This theory predicted an almost scale-invariant spectrum
of cosmological fluctuations, a prediction which has now been
spectacularly confirmed in recent observations \cite{COBE,WMAP}.

Inflationary cosmology is based on the idea that there was a period in
the very early universe during which space expanded almost exponentially.
How to obtain a period of inflation based on fundamental physics has,
however, been a more difficult question to address. In order to obtain
a period of accelerated expansion in the context of Einstein's theory
of General Relativity, the dominant component of matter needs to have
an equation of state with sufficiently negative pressure $p$. More
specifically, the inequality $p < - 1/3 \rho$ (where $\rho$ is the
energy density) is required. The standard approach to obtain inflation
is to invoke the existence of a new scalar field $\varphi$,
the {\it inflaton}, a scalar field
which has to be very weakly coupled to the standard model particle
physics sector. If the energy density of the scalar field is dominated
by its potential energy, then a period of inflation will result.

At this point, the difficulty is shifted to the problem of obtaining
a situation where the energy of a scalar field is dominated by the
potential energy term for a sufficiently long period. The initial
scenario of inflation \cite{Guth,Sato}, in which the scalar field
was assumed to be trapped in a false vacuum, did not work since 
the period of inflation terminated by bubble nucleation, and each
post-inflationary bubble was too small to produce the present
universe (see e.g. \cite{RHBrev,Olive,LL} for reviews of inflationary
cosmology). The next models of inflation 
(``new inflation'' \cite{AS,Linde1}) required initial conditions for
the fields which had to be carefully tuned \cite{MUW,Goldwirth}. 
This led to the development of inflationary models like chaotic
inflation \cite{chaotic} and hybrid inflation \cite{hybrid}, where
inflation happens due to the slow rolling of a field $\phi$ from
initially large values of $|\phi|$ (larger than the Planck scale
in the case of chaotic inflation, smaller in the case of hybrid
inflation, a scenario which invokes the existence of more than
one scalar field). These latter models are much less sensitive to
initial conditions, as shown in \cite{Kung,Feldman}.

As will be argued in the following section, whereas scalar field-driven
inflationary cosmology (in the context of four space-time dimensional
physics using no tools beyond ordinary quantum field theory)
is a very successful scenario, it 
suffers from serious conceptual problems and thus cannot provide
the final theory of the early universe. 
Since several of the key conceptual problems discussed below
relate to ultraviolet issues, it is likely that the same new
fundamental physics required to address the ultraviolet problems
of the Standard Model (SM) of particle physics will be needed to develop a 
true ``theory'' of the very early universe. String theory is
currently our best candidate for resolving the ultraviolet problems
of the particle physics SM, and thus string theory may also provide
us with the theory of the primordial universe, a theory which may
well lead to a period of cosmological inflation - although
the possible existence of alternative scenarios should not be
discarded. 

In this talk, I first discuss some of the conceptual problems of
conventional inflationary cosmology. Then, I formulate some key challenges for
any approach to string cosmology. ``String Gas Cosmology'' (SGC),
a particular approach to combining string theory and cosmology
first put forwards in \cite{BV}, will be briefly reviewed in 
Section 4, and I will conclude with a summary of recent progress in and
current problems of SGC. This article is the first of three
reviews. The second \cite{Review2} will focus on the principles
of SGC, while the third \cite{Review3} will concentrate on recent
progress on the issue of moduli stabilization in SGC.

\section{Conceptual Problems of Inflationary Cosmology}

Before discussing some key conceptual problems of 
conventional scalar field-driven
inflationary cosmology, let us recall some of the key features of
cosmological inflation. To set our notation, we use the following
metric for the homogeneous and isotropic background space-time:
\begin{equation}
ds^2 \, = \, dt^2 - a(t)^2 d{\bf x}^2 \, ,
\end{equation}
where $d{\bf x}^2$ is the metric of ${\cal R}^3$ (we assume for
simplicity a spatially flat universe), and $a(t)$ is the scale factor 
of the universe,

Figure 1 is a sketch of the space-time
structure of an inflationary universe. The vertical axis is time,
the horizontal axis is physical length. The time period between
$t_i$ and $t_R$ is the period of inflation (here for simplicity taken
to be exponential). During the period of inflation, the Hubble radius
$l_H(t) \, \equiv \, H^{-1}(t)$, where $H(t) \, \equiv \, {\dot a(t)} / a(t)$,
is constant. After inflation,
the Hubble radius increases linearly in time. In contrast, the
physical length corresponding to a fixed co-moving scale increases
exponentially during the period of inflation, and then grows either
as $t^{1/2}$ (radiation-dominated phase) or $t^{2/3}$ (matter-dominated
phase), i.e. less fast than the Hubble radius. 
\begin{figure}
\centering
\includegraphics[height=6cm]{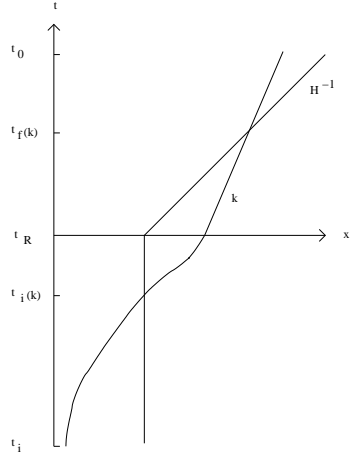}
\caption{Space-time diagram (sketch) showing the evolution
of scales in inflationary cosmology. The vertical axis is
time, and the period of inflation lasts between $t_i$ and
$t_R$, and is followed by the radiation-dominated phase
of standard big bang cosmology. During exponential inflation,
the Hubble radius $H^{-1}$ is constant in physical spatial coordinates
(the horizontal axis), whereas it increases linearly in time
after $t_R$. The physical length corresponding to a fixed
comoving length scale labelled by its wavenumber $k$ increases
exponentially during inflation but increases less fast than
the Hubble radius (namely as $t^{1/2}$), after inflation.}
\label{fig:1}       
\end{figure}

The key feature of inflationary cosmology which can be seen from
Figure 1 is the fact that fixed comoving scales are red-shifted
exponentially relative to the Hubble radius during the period of
inflation. Provided that the period of inflation lasted more than
about $50$ Hubble expansion times (this number is obtained assuming
that the energy scale of inflation is of the order of $10^{16}$GeV),
then modes with a current wavelength up to the Hubble radius started
out at the beginning of the period of inflation with a wavelength smaller
than the Hubble radius at that time. Thus, it is possible to imagine
a microscopic mechanism for creating the density fluctuations in the
early universe which evolve into the cosmological structures we observe
today.

Since during the period of inflation any pre-existing ordinary matter
fluctuations are red-shifted, it is reasonable to assume that quantum
vacuum fluctuations are the source of the currently observed structures
\cite{ChibMukh,Lukash} (see also \cite{Press}). The time-translational
symmetry of the inflationary phase leads, independent of a precise
understanding of the generation mechanism for the fluctuations, to
the prediction that the spectrum of cosmological perturbations should
be approximately scale-invariant \cite{Press,Sato}. 

The quantum theory of linearized cosmological perturbations 
\cite{Sasaki,Mukh}, in particular applied to inflationary
cosmology, has in the mean time become a well-developed research area (see 
e.g. \cite{MFB} for a detailed review, and \cite{RHBrev2} for a
pedagogical introduction). For simple scalar field matter, there is
a single canonically normalized variable, often denoted by $v$, 
which carries the information
about the ``scalar metric fluctuations'', the part of the metric
perturbations which couples at linearized level to the matter.
The equation of motion for each Fourier mode of this variable $v$ has the 
form of a harmonic oscillator with a time-dependent square mass
$m^2$, whose form is set by the cosmological background. On
scales smaller than the Hubble radius, the modes oscillate (quantum
vacuum oscillations). However, on
length scales larger than the Hubble radius, $m^2$ is negative, the
oscillations cease, and
the wave functions of these modes undergo squeezing. Since the squeezing
angle in phase space does not depend on the wave number, all modes
re-enter the Hubble radius at late times with the same squeezing angle.
This then leads to the prediction of ``acoustic'' oscillations in the
angular power spectrum of CMB anisotropies (see e.g.
\cite{acoustic} for a recent analytical treatment), a prediction
spectacularly confirmed by the WMAP data \cite{WMAP}, and allowing
cosmologists to fit for several important cosmological parameters.

In spite of this spectacular phenomenological success of the
inflationary paradigm, I will now argue that conventional scalar field-driven
inflation suffers from several important conceptual problems.

The first problem (the {\bf amplitude problem})
relates to the amplitude of the spectrum of
cosmological perturbations. In a wide class of inflationary
models, obtaining the correct amplitude requires the introduction
of a hierarchy in scales, namely \cite{Adams}
\be
{{V(\varphi)} \over {\Delta \varphi^4}}
\, \leq \, 10^{-12} \, ,
\ee
where $\Delta \varphi$ is the change in the inflaton field during
one Hubble expansion time (during inflation), and $V(\varphi)$ is
the potential energy during inflation.

A more serious problem is the {\bf trans-Planckian problem} \cite{RHBrev3}.
Returning to the space-time diagram of Figure 1, we can immediately
deduce that, provided that the period of inflation lasted sufficiently
long (for GUT scale inflation the number is about 70 e-foldings),
then all scales inside of the Hubble radius today started out with a
physical wavelength smaller than the Planck scale at the beginning of
inflation. Now, the theory of cosmological perturbations is based
on Einstein's theory of General Relativity coupled to a simple
semi-classical description of matter. It is clear that these
building blocks of the theory are inapplicable on scales comparable
and smaller than the Planck scale. Thus, the key
successful prediction of inflation (the theory of the origin of
fluctuations) is based on suspect calculations since 
new physics {\it must} enter
into a correct computation of the spectrum of cosmological perturbations.
The key question is as to whether the predictions obtained using
the current theory are sensitive to the specifics of the unknown
theory which takes over on small scales.
 
One approach to study the sensitivity of the usual predictions of
inflationary cosmology to the unknown physics on trans-Planckian scales
is to study toy models of ultraviolet physics which allow explicit
calculations. The first approach which was used \cite{Jerome1,Niemeyer}
is to replace the usual linear dispersion relation for the Fourier
modes of the fluctuations by a modified dispersion relation, a
dispersion relation which is linear for physical wavenumbers smaller
than the scale of new physics, but deviates on larger scales. Such
dispersion relations were used previously to test the sensitivity
of black hole radiation on the unknown physics of the UV 
\cite{Unruh,CJ}. It was found \cite{Jerome1} that if the evolution of modes on
the trans-Planckian scales is non-adiabatic, then substantial
deviations of the spectrum of fluctuations from the usual results
are possible. Non-adiabatic evolution turns an initial state
minimizing the energy density into a state which is excited once
the wavelength becomes larger than the cutoff scale. Back-reaction
effects of these excitations may limit the magnitude of the
trans-Planckian effects, but - based on our recent study \cite{Jerome2} -
not to the extent initially assumed \cite{Tanaka,Starob3}.
Other approaches to study the trans-Planckian problem have been 
pursued, e.g. based on implementing the space-space \cite{Easther}
or space-time \cite{Ho} uncertainty relations, on a minimal length
hypothesis \cite{Kempf}, on ``minimal trans-Planckian'' assumptions (taking
as initial conditions some vacuum state at the mode-dependent time
when the wavelength of the mode is equal to the Planck scale
\cite{minimal}, or on effective field theory \cite{Cliff}, all showing
the possibility of trans-Planckian corrections.

A third problem is the {\bf singularity problem}. It was known for a long
time that standard Big Bang cosmology cannot be the complete story of
the early universe because of
the initial singularity, a singularity which is unavoidable when basing
cosmology on Einstein's field equations in the presence of a matter
source obeying the weak energy conditions (see e.g. \cite{HE} for
a textbook discussion). Recently, the singularity theorems have been
generalized to apply to Einstein gravity coupled to scalar field
matter, i.e. to scalar field-driven inflationary cosmology \cite{Borde}.
It is shown that in this context, a past singularity at some point
in space is unavoidable. Thus we know, from the outset, that scalar
field-driven inflation cannot be the ultimate theory of the very
early universe.

The Achilles heel of scalar field-driven inflationary cosmology is,
however, the {\bf cosmological constant problem}. We know from
observations that the large quantum vacuum energy of field theories
does not gravitate today. However, to obtain a period of inflation
one is using the part of the energy-momentum tensor of the scalar field
which looks like the vacuum energy. In the absence of a convincing
solution of the cosmological constant problem it is unclear whether
scalar field-driven inflation is robust, i.e. whether the
mechanism which renders the quantum vacuum energy gravitationally
inert today will not also prevent the vacuum energy from
gravitating during the period of slow-rolling of the inflaton 
field. Note that the
approach to addressing the cosmological constant problem making use
of the gravitational back-reaction of long range fluctuations
(see \cite{RHBrev4} for a summary of this approach) does not prevent
a long period of inflation in the early universe.

Finally, a key challenge for conventional inflationary cosmology is to find a
well-motivated candidate for the scalar field which drives inflation,
the inflaton. Ever since the failure of the model of {\it old inflation}
\cite{Guth,Sato}, it is clear that physics beyond the Standard Model
of particle physics must be invoked. 

\section{Challenges for String Cosmology}

In the following we will focus not on the question of how to obtain
inflation from a new fundamental theory of micro-physics, but rather
on how such new micro-physics can help resolve some of the key conceptual
problems of scalar field-driven inflation listed in the previous
section. To be specific, we will assume that the new fundamental
physics is based on superstring theory.

Why could superstring theory help resolve the key problems mentioned
in the previous section? First, an effective field theory derived from
string theory contains many scalar fields which are massless before
supersymmetry breaking. Thus, the hierarchy of scales required to produce
the observed small amplitude of cosmological perturbations might arise
naturally. Second, string theory is supposed to provide a theory
which describes physics on all scales. In such a theory, it would be
possible to track the cosmological perturbations for all times. We would
know the trans-Planckian physics and would thus be able to compute
the trans-Planckian signatures. Thirdly, one of the goals of string
theory is to resolve all singularities. If this goal can be achieved
(and a scenario which can eliminate cosmological singularities is
discussed below), then it would be possible to construct a theory of
the early universe which does not suffer from the singularity problem
of scalar field-driven inflation. At the moment, it is not clear
whether string theory can address the cosmological constant problem.
It is possible, however, that the cosmological constant problem can
be cured by a better understanding of the infrared physics in the
context of our existing micro-physical theory (see \cite{RHBrev4}
for a review and \cite{woodard} for a key original work).

An immediate problem which arises when trying to connect string theory
with cosmology is the {\bf dimensionality problem}. Superstring theory
is perturbatively consistent only in 10 space-time dimensions, but we
only see 3 large spatial dimensions. The original approach to 
addressing this problem is to assume that the 6 extra dimensions are
compactified on a very small space which cannot be probed with our
available energies. However, from the point of view of cosmology,
it is quite unsatisfactory not to be able to understand why it is
precisely 3 dimensions which are not compactified and why the compact
dimensions are stable. Brane world cosmology \cite{brane} provides
another approach to this problem: it assumes that we live on a
three-dimensional brane embedded in a large nine-dimensional space.
Once again, a cosmologically satisfactory theory should explain
why it is likely that we will end up exactly on a three-dimensional
brane (for some interesting work addressing this issue see
\cite{Mahbub,Mairi,Lisa}).

Finding a natural solution to the dimensionality problem is thus one
of the key challenges for superstring cosmology. This challenge has
various aspects. First, there must be a mechanism which singles out
three dimensions as the number of spatial dimensions we live in.
Second, the moduli fields which describe the volume and the shape of
the unobserved dimensions must be stabilized (any strong time-dependence
of these fields would lead to serious phenomenological constraints).
This is the {\bf moduli problem} for superstring cosmology. As
mentioned above, resolving the {\bf singularity problem} is another of
the main challenges. These are the three problems which {\bf string
gas cosmology} \cite{BV,TV,ABE} explicitly addresses at the present
level of development..

In order to make successful connection with late time cosmology,
any approach to string cosmology must also solve the 
{\bf flatness problem}, namely make sure that the three large
spatial dimensions obtain a sufficiently high entropy (size) to
explain the current universe. Finally, it must provide a
mechanism to produce a nearly scale-invariant spectrum of
nearly adiabatic cosmological perturbations.

Since superstring theory leads to many light scalar fields, it is
 possible that superstring cosmology will provide a
convincing realization of inflation (see e.g. \cite{stringinflation}
for reviews of recent work attempting to obtain inflation in the
context of string theory). However, it is also possible that
superstring cosmology will provide an alternative to cosmological
inflation, maybe along the lines of the Pre-Big-Bang \cite{PBB}
or Ekpyrotic \cite{KOST} scenarios. The greatest challenge for
these alternatives is to solve the flatness problem (see e.g.
\cite{Pyro}). 

\section{Overview of String Gas Cosmology}

A key obstacle towards making progress in developing superstring
cosmology is that non-perturbative string theory does not yet
exist. It is likely that the new structures of non-perturbative
string theory will reveal quite new and unexpected possibilities for
cosmology in a similar way that the non-perturbative understanding
of quantum particle dynamics was crucial in developing inflationary cosmology.

In the absence of a non-perturbative formulation of string theory,
the approach to string cosmology which we have suggested \cite{BV,TV,ABE}
(see also \cite{Perlt})
is to focus on symmetries and degrees of freedom which are new to
string theory (compared to point particle theories) and which will
be part of a non-perturbative string theory, and to use
them to develop a new cosmology. The symmetry we make use of is
{\bf T-duality}, and the new degrees of freedom are 
{\bf string winding modes}.

Let us assume that all spatial directions are toroidal, with
$R$ denoting the radius of the torus. Strings have three types
of states: {\it momentum modes} which represent the center
of mass motion of the string, {\it oscillatory modes} which
represent the fluctuations of the strings, and {\it winding
modes} counting the number of times a string wraps the torus.
Both oscillatory and winding states are special to strings as
opposed to point particles. 

The energy of an oscillatory mode is independent of $R$, momentum
mode energies are quantized in units of $1/R$, i.e.
\be
E_n \, = \, n {1 \over R} \, ,
\ee
and winding mode energies are quantized in units of $R$:
\be
E_m \, = \, m R \, ,
\ee
where both $n$ and $m$ are integers.

The T-duality symmetry is a symmetry of the spectrum of string
states under the change
\be \label{Tdual}
R \, \rightarrow \, {1 \over R}
\ee
in the radius of the torus (in units of the string length $l_s$).
Under such a change, the energy spectrum of string states is
invariant: together with the transformation (\ref{Tdual}), winding
and momentum quantum numbers need to be interchanged
\be \label{Tdual2}
(n, m) \, \rightarrow \, (m, n) \, .
\ee
The string vertex operators are consistent with this symmetry, and
thus T-duality is a symmetry of perturbative string theory. Postulating
that T-duality extends to non-perturbative string theory leads
\cite{Pol} to the need of adding D-branes to the list of fundamental
objects in string theory. With this addition, T-duality is expected
to be a symmetry of non-perturbative string theory.
 
One deficiency inherent in any approach to superstring cosmology
in the absence of a non-perturbative formulation of string theory
is the requirement to introduce a classical background. We choose
the background to be dilaton gravity. It is crucial to include the
dilaton in the Lagrangian, firstly since
the dilaton arises in string perturbation theory at the same level
as the graviton, and secondly because it is only the action of
dilaton gravity (rather than the action of Einstein gravity)
which is consistent with the T-duality symmetry. Given this
background, we consider an ideal gas of matter made up of all
fundamental states of string theory, in particular including
string winding modes.
 
Any physical theory requires initial conditions. We assume that
the universe starts out small and hot. For simplicity, we take
space to be toroidal, with radii in all spatial directions given by
the string scale. We assume that the initial energy density 
is very high, with an effective temperature which is close
to the Hagedorn temperature \cite{Hagedorn}, the maximal temperature
of perturbative string theory. It is not important for us to
have perfect thermal equilibrium, but we need to assume that all
string states (including states with winding and momentum) are
initially excited.

The first predictions of string gas cosmology (SGC) were reached by
heuristic considerations \cite{BV}. Based on the T-duality
symmetry, it was argued that the cosmology resulting from SGC
will be non-singular. For example, as the background radius $R$
varies, the physical temperature $T$ will obey the symmetry
\be
T(R) \, = \, T(1/R)
\ee
and thus remain non-singular even if $R$ decreases to zero.
Similarly, the length $L$ measured by a physical observer will
be consistent with the symmetry (\ref{Tdual}), hence realizing
the idea of a minimal physical length.

Next, it was argued \cite{BV} that in order for spatial sections to become
large, the winding modes need to decay. This decay, at least
on a background with stable one cycles such as a torus, is only
possible if two winding modes meet and annihilate. Since string
world sheets have measure zero probability for intersecting in more
than four space-time dimensions, winding modes can annihilate only
in three spatial dimensions (see, however, the recent
caveats to this conclusion based on the work of \cite{Kabat3}). 
Thus, only three spatial dimensions
can become large, hence explaining the observed dimensionality of
space-time. As was shown later \cite{ABE}, adding branes to
the system does not change these conclusions since at later
times the strings dominate the cosmological dynamics.
Note that in the three dimensions which are becoming large there
is a natural mechanism of isotropization as long as some winding
modes persist \cite{Watson1}.

The equations of SGC are based on coupling an ideal gas of all
string and brane modes to the background space-time of
dilaton gravity. These equations were first studied in
\cite{TV} (see also \cite{Ven}). In the context of a
homogeneous and isotropic background metric, it follows,
as will be reviewed in \cite{Review2}, that - in the
absence of string interactions - the scale factor remains bounded
from above and below. String winding modes resist the expansion,
string momentum modes resist the contraction. Given a gas of
strings with equal numbers of winding and momentum modes, the
equilibrium state for the radius is the string scale (the ``self-dual''
radius). Note that the dilaton is not stabilized at this stage.

The interaction of string winding modes results in string loop
production. This process was studied in the three large dimensions
in \cite{BEK}, with the conclusion that, after a stage of ``loitering'',
the three spatial dimensions are liberated and expand, as long as
the dilaton is not too small (this last caveat follows from the
work of \cite{Kabat3}).

\section{Progress and Problems in String Gas Cosmology}

A key issue in all approaches to string cosmology is the question of
{\bf moduli stabilization}. The challenge is to fix the shape and
volume moduli of the compact dimensions and to fix the value of
the dilaton. Moduli stabilization is essential to obtain a 
consistent late time cosmology.

There has recently been a lot of progress on the issue of moduli
stabilization in SGC, progress which will be reviewed in
detail in \cite{Review3}. In a first study \cite{Watson2}, the stabilization
of the radii of the extra dimensions (the ``radion'' degrees of freedom)
was studied in the string frame. It was shown that, as long as there are
an equal number of string momentum and winding modes about the
compact directions, the radii are dynamically stabilized at the 
self-dual radius. The dilaton, however, is in general evolving in time.

For late time cosmology, it is crucial to show that the radion degrees
of freedom are stabilized in the Einstein frame. Obstacles towards
achieving this goal were put forward in \cite{BattWat,Aaron,Tirtho,Damien}.
However, if the spectrum of string states contains modes which are
massless at the self-dual radius (which is the case for heterotic
but not for Type II string theory), then these modes generate an
effective potential for the radion which has a minimum with vanishing
energy at the self-dual radius and thus yield radion stabilization
\cite{Subodh1,Subodh2} (see also \cite{Watson3}). 
As shown in these references, the
radion stabilization mechanism is consistent with late time
cosmology (e.g. fifth force constraints).
These same massless modes also yield  stabilization
of the shape moduli \cite{Edna} (see also \cite{Sugumi}
for a study which appeared after the conference). 
The outstanding challenge in this
approach is to stabilize the dilaton (for some ideas see \cite{Subodh3}). 

There has been a substantial amount of recent work on SGC. For lack of
space, references to this work will be given in \cite{Review2}. Although
the moduli stabilization mechanism in SGC is conceptually simpler
than the mechanisms used in other approaches to string cosmology
(see e.g. \cite{GKP}) in that it does not use fluxes or warping of
the internal dimensions, that is uses more string-specific ingredients,
and that it is based on a full dynamical analysis, the applicability of
the mechanism at first sight depends on special topological features
of the internal space, namely on the existence of stable one cycles.
However, as shown in \cite{Kabat1}, the basic mechanisms of SGC
generalize to certain orbifolds. They may also be generalized using
ideas of \cite{Stephon} to more general Calabi-Yau spaces.

A crucial challenge for SGC is to find a solution to the flatness
or entropy problem, namely to produce a sufficiently large 
three-dimensional universe to explain its observed size. The only
avenue to achieve this which is well-established at the present
time is by invoking a period of inflation of
the three large dimensions at later times. For some ideas on
how to obtain inflation from SGC see \cite{Moshe,Easson2,Tirtho2}.
A challenge for inflation in the context of SGC is to ensure that
inflation is compatible with radion stabilization. Note, however,
that since SGC naturally seems to point to a bouncing cosmology,
eventually an alternative to inflation in the context of SGC may
emerge. In this case, however, the challenge would be to find an
alternative mechanism for generating a nearly scale-invariant
spectrum of nearly adiabatic cosmological fluctuations.

\centerline{\bf Acknowledgements}

I would like to thank Manu Paranjape and Richard MacKenzie for
organizing Theory Canada 1, for their hospitality during the
meeting, and for inviting me to speak. This research is
supported by an NSERC Discovery Grant and by the Canada Research
Chairs program.


\begin{thebibliography}{9}

\bibitem{Guth}
A.~H.~Guth,
  Phys.\ Rev.\ D {\bf 23}, 347 (1981).

\bibitem{Sato}
K.~Sato,
  Mon.\ Not.\ Roy.\ Astron.\ Soc.\  {\bf 195}, 467 (1981).

\bibitem{Starob1}
A.~A.~Starobinsky,
  Phys.\ Lett.\ B {\bf 91}, 99 (1980).

\bibitem{Brout}
R.~Brout, F.~Englert and E.~Gunzig,
  Annals Phys.\  {\bf 115}, 78 (1978).

\bibitem{ChibMukh}
V.~F.~Mukhanov and G.~V.~Chibisov,
JETP Lett.\  {\bf 33}, 532 (1981)
[Pisma Zh.\ Eksp.\ Teor.\ Fiz.\  {\bf 33}, 549 (1981)].

\bibitem{Lukash}
V.~N.~Lukash, Pisma Zh. Eksp. Teor. Fiz. {\bf 31}, 631 (1980);\\
V.~N.~Lukash,
Sov.\ Phys.\ JETP {\bf 52}, 807 (1980)
[Zh.\ Eksp.\ Teor.\ Fiz.\  {\bf 79},  (1980)].

\bibitem{Press}
W. Press, Phys. Scr. {\bf 21}, 702 (1980).

\bibitem{Starob2}
A.~A.~Starobinsky,
  JETP Lett.\  {\bf 30}, 682 (1979)
  [Pisma Zh.\ Eksp.\ Teor.\ Fiz.\  {\bf 30}, 719 (1979)].

\bibitem{COBE}
G.~F.~Smoot {\it et al.},
  Astrophys.\ J.\  {\bf 396}, L1 (1992).

\bibitem{WMAP}
C.~L.~Bennett {\it et al.},
  Astrophys.\ J.\ Suppl.\  {\bf 148}, 1 (2003)
  [arXiv:astro-ph/0302207].

\bibitem{RHBrev}
R.~H.~Brandenberger,
  Rev.\ Mod.\ Phys.\  {\bf 57}, 1 (1985).

\bibitem{Olive}
K.~A.~Olive,
  Phys.\ Rept.\  {\bf 190}, 307 (1990).

\bibitem{Lindebook}
A. Linde, \textit{Particle Physics and Inflationary
Cosmology}, (Harwood, Chur, 1990).

\bibitem{LL}
A.~R.~Liddle and D.~H.~Lyth,
\textit{Cosmological inflation and large-scale structure},
(Cambridge Univ. Press, Cambridge, 2000).

\bibitem{AS}
A.~Albrecht and P.~J.~Steinhardt,
  Phys.\ Rev.\ Lett.\  {\bf 48}, 1220 (1982).

\bibitem{Linde1}
A.~D.~Linde,
  Phys.\ Lett.\ B {\bf 108}, 389 (1982).

\bibitem{MUW}
G.~F.~Mazenko, R.~M.~Wald and W.~G.~Unruh,
  Phys.\ Rev.\ D {\bf 31}, 273 (1985).

\bibitem{Goldwirth}
D.~S.~Goldwirth and T.~Piran,
  Phys.\ Rept.\  {\bf 214}, 223 (1992).

\bibitem{chaotic}
A.~D.~Linde,
  Phys.\ Lett.\ B {\bf 129}, 177 (1983).

\bibitem{hybrid}
A.~D.~Linde,
  Phys.\ Rev.\ D {\bf 49}, 748 (1994)
  [arXiv:astro-ph/9307002].

\bibitem{Kung}
R.~H.~Brandenberger and J.~H.~Kung,
  Phys.\ Rev.\ D {\bf 42}, 1008 (1990).

\bibitem{Feldman}
H.~A.~Feldman and R.~H.~Brandenberger,
  Phys.\ Lett.\ B {\bf 227}, 359 (1989).

\bibitem{BV}
R.~H.~Brandenberger and C.~Vafa,
  Nucl.\ Phys.\ B {\bf 316}, 391 (1989).

\bibitem{Review2}
R. Brandenberger, [arXiv:hep-th/0509099]
\textit{Challenges for String Gas Cosmology},
to appear in the proceedings of the 59th Yamada conference
(Univ. of Tokyo, Tokyo, Japan, June 20 - 24, 2005).

\bibitem{Review3}
R. Brandenberger, [arXiv:hep-th/0509159]
\textit{Moduli Stabilization in String Gas Cosmology}
to appear in the proceedings of YKIS 2005 (Yukawa Institute for
Theoretical Physics, Kyoto, Japan, June 27 - July 1, 2005).

\bibitem{Sasaki}
M.~Sasaki,
  Prog.\ Theor.\ Phys.\  {\bf 76}, 1036 (1986).

\bibitem{Mukh}
V.~F.~Mukhanov,
  Sov.\ Phys.\ JETP {\bf 67}, 1297 (1988)
  [Zh.\ Eksp.\ Teor.\ Fiz.\  {\bf 94N7}, 1 (1988)].

\bibitem{MFB}
V.~F.~Mukhanov, H.~A.~Feldman and R.~H.~Brandenberger,
Phys.~Rept.~{\bf 215}, 203 (1992);\\
 J.~Martin, [arXiv:hep-th/0406011].

\bibitem{RHBrev2}
R.~H.~Brandenberger,
  Lect.\ Notes Phys.\  {\bf 646}, 127 (2004)
  [arXiv:hep-th/0306071].

\bibitem{acoustic}
V.~Mukhanov,
  arXiv:astro-ph/0303072.

\bibitem{Adams}
F.~C.~Adams, K.~Freese and A.~H.~Guth,
  Phys.\ Rev.\ D {\bf 43}, 965 (1991).

\bibitem{RHBrev3}
R.~H.~Brandenberger, [arXiv:hep-ph/9910410].

\bibitem{Jerome1}
R.~H.~Brandenberger and J.~Martin, Mod.~Phys.~Lett.~A~{\bf 16}, 999
(2001), [arXiv:astro-ph/0005432];\\
J.~Martin and R.~H.~Brandenberger,
Phys.~Rev.~D~{\bf 63}, 123501 (2001), [arXiv:hep-th/0005209].

\bibitem{Niemeyer}
J.~C.~Niemeyer, Phys.~Rev.~D~{\bf 63}, 123502 (2001),
[arXiv:astro-ph/0005533]; \\
S.~Shankaranarayanan, Class.~Quant.~Grav.~{\bf 20}, 75 (2003), 
[arXiv:gr-qc/0203060];\\
J.~C.~Niemeyer and R.~Parentani, Phys.~Rev.~D~{\bf 64}, 101301 (2001),
[arXiv:astro-ph/0101451].

\bibitem{Unruh}
W.~G.~Unruh,
  Phys.\ Rev.\ D {\bf 51}, 2827 (1995).

\bibitem{CJ}
S.~Corley and T.~Jacobson,
  Phys.\ Rev.\ D {\bf 54}, 1568 (1996)
  [arXiv:hep-th/9601073].

\bibitem{Jerome2}
R.~H.~Brandenberger and J.~Martin,
  Phys.\ Rev.\ D {\bf 71}, 023504 (2005)
  [arXiv:hep-th/0410223];\\
B.~R.~Greene, K.~Schalm, G.~Shiu and J.~P.~van der Schaar,
  JCAP {\bf 0502}, 001 (2005)
  [arXiv:hep-th/0411217];\\
U.~H.~Danielsson,
  Phys.\ Rev.\ D {\bf 71}, 023516 (2005)
  [arXiv:hep-th/0411172].

\bibitem{Tanaka}
T.~Tanaka, [arXiv:astro-ph/0012431].

\bibitem{Starob3}
A.~A.~Starobinsky, Pisma Zh.~Eksp.~Teor.~Fiz.~{\bf 73}, 415 (2001),
[JETP Lett.\ {\bf 73}, 371 (2001)], [arXiv:astro-ph/0104043].

\bibitem{Easther}
C.~S.~Chu, B.~R.~Greene and G.~Shiu, Mod.~Phys.~Lett.~A~{\bf 16}, 2231
(2001), [arXiv:hep-th/0011241]; \\
R.~Easther, B.~R.~Greene, W.~H.~Kinney
and G.~Shiu, Phys.~Rev.~D~{\bf 64}, 103502 (2001),
[arXiv:hep-th/0104102]; \\
R.~Easther, B.~R.~Greene, W.~H.~Kinney and
G.~Shiu, Phys.~Rev.~D~{\bf 67}, 063508 (2003), [arXiv:hep-th/0110226];\\
F.~Lizzi, G.~Mangano, G.~Miele and
M.~Peloso, JHEP~{\bf 0206}, 049 (2002) [arXiv:hep-th/0203099];\\
S.~F.~Hassan and M.~S.~Sloth, Nucl.~Phys.~B~{\bf 674}, 434 (2003),
[arXiv:hep-th/0204110].

\bibitem{Ho}
R.~Brandenberger and P.~M.~Ho, Phys.~Rev.~D~{\bf 66}, 023517 (2002),
[AAPPS Bull.~{\bf 12N1}, 10 (2002)], [arXiv:hep-th/0203119].

\bibitem{Kempf}
A.~Kempf and J.~C.~Niemeyer, Phys.~Rev.~D~{\bf 64}, 103501 (2001),
[arXiv:astro-ph/0103225].

\bibitem{minimal}
U.~H.~Danielsson, Phys.~Rev.~D~{\bf 66}, 023511 (2002),
[arXiv:hep-th/0203198];\\
V.~Bozza, M.~Giovannini and G.~Veneziano, JCAP~{\bf 0305}, 001 (2003),
[arXiv:hep-th/0302184];\\
 J.~C.~Niemeyer, R.~Parentani and D.~Campo,
Phys.~Rev.~D~{\bf 66}, 083510 (2002), [arXiv:hep-th/0206149].

\bibitem{Cliff}
C.~P.~Burgess, J.~M.~Cline, F.~Lemieux and R.~Holman,
  JHEP {\bf 0302}, 048 (2003)
  [arXiv:hep-th/0210233];\\
K.~Schalm, G.~Shiu and J.~P.~van der Schaar,
  AIP Conf.\ Proc.\  {\bf 743}, 362 (2005)
  [arXiv:hep-th/0412288].

\bibitem{HE}
S. Hawking and G. Ellis, \textit{The Large-Scale Structure of Space-Time}
(Cambridge Univ. Press, Cambridge, 1973).

\bibitem{Borde}
A.~Borde and A.~Vilenkin,
  Phys.\ Rev.\ Lett.\  {\bf 72}, 3305 (1994)
  [arXiv:gr-qc/9312022].

\bibitem{RHBrev4}
R.~H.~Brandenberger,
Plenary talk at 18th IAP Colloquium on the Nature of Dark Energy: 
Observational and Theoretical Results on the Accelerating Universe, 
Paris, France, 1-5 Jul 2002.
arXiv:hep-th/0210165.

\bibitem{woodard}
N.~C.~Tsamis and R.~P.~Woodard,
Phys.\ Lett.\ B {\bf 301}, 351 (1993);\\
N.~C.~Tsamis and R.~P.~Woodard,
Nucl.\ Phys.\ B {\bf 474}, 235 (1996)
[arXiv:hep-ph/9602315].

\bibitem{brane}
L.~Randall and R.~Sundrum,
  Phys.\ Rev.\ Lett.\  {\bf 83}, 4690 (1999)
  [arXiv:hep-th/9906064].

\bibitem{Mahbub}
M.~Majumdar and A.-C. Davis,
  JHEP {\bf 0203}, 056 (2002)
  [arXiv:hep-th/0202148].

\bibitem{Mairi}
R.~Durrer, M.~Kunz and M.~Sakellariadou,
  Phys.\ Lett.\ B {\bf 614}, 125 (2005)
  [arXiv:hep-th/0501163].

\bibitem{Lisa}
A.~Karch and L.~Randall,
  arXiv:hep-th/0506053.

\bibitem{TV}
A.~A.~Tseytlin and C.~Vafa,
  Nucl.\ Phys.\ B {\bf 372}, 443 (1992)
  [arXiv:hep-th/9109048].

\bibitem{ABE}
S.~Alexander, R.~H.~Brandenberger and D.~Easson,
  Phys.\ Rev.\ D {\bf 62}, 103509 (2000)
  [arXiv:hep-th/0005212].

\bibitem{stringinflation}
C.~P.~Burgess,
  Pramana {\bf 63}, 1269 (2004)
  [arXiv:hep-th/0408037];\\
A.~Linde,
  eConf {\bf C040802}, L024 (2004)
  [arXiv:hep-th/0503195];\\
J.~M.~Cline,
  arXiv:hep-th/0501179.

\bibitem{PBB}
M.~Gasperini and G.~Veneziano,
  Astropart.\ Phys.\  {\bf 1}, 317 (1993)
  [arXiv:hep-th/9211021].

\bibitem{KOST}
J.~Khoury, B.~A.~Ovrut, P.~J.~Steinhardt and N.~Turok,
  Phys.\ Rev.\ D {\bf 64}, 123522 (2001)
  [arXiv:hep-th/0103239].

\bibitem{Pyro}
R.~Kallosh, L.~Kofman and A.~D.~Linde,
  Phys.\ Rev.\ D {\bf 64}, 123523 (2001)
  [arXiv:hep-th/0104073].

\bibitem{Perlt}
J.~Kripfganz and H.~Perlt,
  Class.\ Quant.\ Grav.\  {\bf 5}, 453 (1988).

\bibitem{Pol}
J. Polchinski, \textit{String Theory, Vols. 1 and 2},
(Cambridge Univ. Press, Cambridge, 1998).

\bibitem{Hagedorn}
R.~Hagedorn,
  Nuovo Cim.\ Suppl.\  {\bf 3}, 147 (1965).

\bibitem{Kabat3}
R.~Easther, B.~R.~Greene, M.~G.~Jackson and D.~Kabat,
  JCAP {\bf 0502}, 009 (2005)
  [arXiv:hep-th/0409121];\\
R.~Danos, A.~R.~Frey and A.~Mazumdar,
  Phys.\ Rev.\ D {\bf 70}, 106010 (2004)
  [arXiv:hep-th/0409162].

\bibitem{Watson1}
S.~Watson and R.~H.~Brandenberger,
  Phys.\ Rev.\ D {\bf 67}, 043510 (2003)
  [arXiv:hep-th/0207168].

\bibitem{Ven}
G.~Veneziano,
  Phys.\ Lett.\ B {\bf 265}, 287 (1991).

\bibitem{BEK}
R.~Brandenberger, D.~A.~Easson and D.~Kimberly,
  Nucl.\ Phys.\ B {\bf 623}, 421 (2002)
  [arXiv:hep-th/0109165].

\bibitem{Watson2}
S.~Watson and R.~Brandenberger,
  JCAP {\bf 0311}, 008 (2003)
  [arXiv:hep-th/0307044].

\bibitem{BattWat}
T.~Battefeld and S.~Watson,
  JCAP {\bf 0406}, 001 (2004)
  [arXiv:hep-th/0403075].

\bibitem{Aaron}
A.~J.~Berndsen and J.~M.~Cline,
  Int.\ J.\ Mod.\ Phys.\ A {\bf 19}, 5311 (2004)
  [arXiv:hep-th/0408185].

\bibitem{Tirtho}
A.~Berndsen, T.~Biswas and J.~M.~Cline,
  arXiv:hep-th/0505151.

\bibitem{Damien}
D.~A.~Easson and M.~Trodden,
  Phys.\ Rev.\ D {\bf 72}, 026002 (2005)
  [arXiv:hep-th/0505098].

\bibitem{Subodh1}
S.~P.~Patil and R.~Brandenberger,
  Phys.\ Rev.\ D {\bf 71}, 103522 (2005)
  [arXiv:hep-th/0401037].

\bibitem{Subodh2}
S.~P.~Patil and R.~H.~Brandenberger,
  arXiv:hep-th/0502069.

\bibitem{Watson3}
S.~Watson,
  Phys.\ Rev.\ D {\bf 70}, 066005 (2004)
  [arXiv:hep-th/0404177].

\bibitem{Edna}
R.~Brandenberger, Y.~K.~Cheung and S.~Watson,
  arXiv:hep-th/0501032.

\bibitem{Sugumi}
S. Kanno and J. Soda,
  arXiv:hep-th/0509074.

\bibitem{Subodh3}
S.~P.~Patil,
  arXiv:hep-th/0504145.

\bibitem{GKP}
S.~B.~Giddings, S.~Kachru and J.~Polchinski,
  Phys.\ Rev.\ D {\bf 66}, 106006 (2002)
  [arXiv:hep-th/0105097].

\bibitem{Kabat1}
R.~Easther, B.~R.~Greene and M.~G.~Jackson,
  Phys.\ Rev.\ D {\bf 66}, 023502 (2002)
  [arXiv:hep-th/0204099].

\bibitem{Stephon}
S.~H.~S.~Alexander,
  JHEP {\bf 0310}, 013 (2003)
  [arXiv:hep-th/0212151].

\bibitem{Moshe}
S.~Alexander, R.~Brandenberger and M.~Rozali,
  arXiv:hep-th/0302160.

\bibitem{Easson2}
R.~Brandenberger, D.~A.~Easson and A.~Mazumdar,
  Phys.\ Rev.\ D {\bf 69}, 083502 (2004)
  [arXiv:hep-th/0307043].

\bibitem{Tirtho2}
T.~Biswas, R.~Brandenberger, D.~A.~Easson and A.~Mazumdar,
  Phys.\ Rev.\ D {\bf 71}, 083514 (2005)
  [arXiv:hep-th/0501194].

\end{thebibliography}
\end{document}